\documentclass[a4paper,12pt]{article}
\usepackage{amssymb,amstext,amsmath,amsthm,mathrsfs,mathtools,bbm}
\pdfoutput=1

\usepackage{setspace} 
\usepackage{enumerate}
\usepackage[table,dvipsnames]{xcolor}
\usepackage{graphicx}
\usepackage{xspace}
\usepackage{booktabs}
\usepackage{tabularx}
\usepackage{lscape}
\usepackage{algorithmic, algorithm}
\usepackage{subfig}

\usepackage{natbib} 
\bibpunct{(}{)}{,}{a}{,}{,}

\usepackage{geometry} 
 \usepackage[noabbrev]{cleveref}
 
\newcommand{\logit}{\textsc{logit}\xspace} 
\newcommand{\ols}{\textsc{ols}\xspace} 
\newcommand{\mabbuckets}{\textsc{b-bucket}\xspace} 
\newcommand{\mabgrid}{\textsc{b-grid}\xspace} 
\newcommand{\mabmodels}{\textsc{b-model}\xspace} 
\newcommand{\ml}{\textsc{ml}\xspace} 
\newcommand{\greedy}{\textsc{greedy}\xspace} 
\newcommand{\wls}{\textsc{wls}\xspace} 
\newcommand*\rot{\rotatebox{90}}

\DeclareMathOperator*{\argmax}{arg\,max}

\newcolumntype{A}{>{\raggedright\arraybackslash\hsize=0.28\hsize}X}
\newcolumntype{B}{>{\raggedright\arraybackslash\hsize=1.34\hsize}X}
\newcolumntype{C}{>{\raggedright\arraybackslash\hsize=1.84\hsize}X}
\newcolumntype{D}{>{\raggedright\arraybackslash\hsize=0.54\hsize}X}

\begin{document}
\begin{center}
\textbf{\Large Dynamic Pricing and Learning with Competition: \\ Insights from the Dynamic Pricing Challenge at the 2017 INFORMS RM \& Pricing Conference\footnote{Chris Bayliss and Christine Currie were funded by the EPSRC under grant number EP/N006461/1. Andria Ellina and Simos Zachariades were part funded by EPSRC as part of their PhD studentships. Asbj\o rn Nilsen Riseth was partially funded by EPSRC grant EP/L015803/1.}\footnote{Corresponding author: Ruben van de Geer, e-mail: r.vande.geer-at-vu.nl, phone: +31(0)644810227.}} \vspace{0.5cm} \\
\today
\end{center}
\begin{flushleft}
\textbf{Ruben van de Geer\textsuperscript{a,I},
Arnoud V.\ den Boer\textsuperscript{b,I},
Christopher Bayliss\textsuperscript{c,II},
Christine Currie\textsuperscript{c,II},
Andria Ellina\textsuperscript{c,II},
Malte Esders\textsuperscript{d,II},
Alwin Haensel\textsuperscript{e,II},
Xiao Lei\textsuperscript{f,II},
Kyle D.S.\ Maclean\textsuperscript{g,II},
Antonio Martinez-Sykora\textsuperscript{c,II},
Asbj\o rn Nilsen Riseth\textsuperscript{h,II},
Fredrik \O degaard\textsuperscript{g,II},
and
Simos Zachariades\textsuperscript{c,II}
} \vspace{0.2cm}\\
{\footnotesize\textit{\textsuperscript{a}Department of Mathematics, Vrije Universiteit, De Boelelaan 1105, 1081 HV Amsterdam, The Netherlands;
\textsuperscript{b}Korteweg-de Vries Institute for Mathematics, University of Amsterdam, Science Park 105-107, 1098 XG Amsterdam, The Netherlands, and
Amsterdam Business School, University of Amsterdam, Plantage Muidergracht 12, 1018 TV Amsterdam, The Netherlands; 
\textsuperscript{c}Mathematical Sciences, University of Southampton, SO17 1BJ, UK; 
\textsuperscript{d}Faculty IV Electrical Engineering and Computer Science, Technische Universit\"at Berlin, Stra{\ss}e des 17. Juni 135,
10623 Berlin;
\textsuperscript{e}Haensel AMS, Advanced Mathematical Solutions, Ritterstr. 2a, 10969 Berlin, Germany;
\textsuperscript{f}Department of Industrial Engineering and Operations Research, Columbia University, New York, NY 10027;
\textsuperscript{g}Ivey Business School, Western University, 1255 Western Road, London, ON N6G 0N1, Canada;
\textsuperscript{h}Mathematical Institute, University of Oxford, OX2 6GG, UK.
}} \vspace{0.2cm}\\
{\footnotesize\textit{\noindent\textsuperscript{I}dynamic pricing challenge organizer, lead author; \\
\textsuperscript{II}dynamic pricing challenge participant, co-author.
}} \vspace{0.2cm}\\
\end{flushleft}
%
\begin{center}
Abstract
\end{center}
This paper presents the results of the Dynamic Pricing Challenge, held on the occasion of the 17\textsuperscript{th} INFORMS Revenue Management and Pricing Section Conference on June 29-30, 2017 in Amsterdam, The Netherlands. For this challenge, participants submitted algorithms for pricing and demand learning of which the numerical performance was analyzed in simulated market environments. This allows consideration of market dynamics that are not analytically tractable or can not be empirically analyzed due to practical complications. Our findings implicate that the relative performance of algorithms varies substantially across different market dynamics, which confirms the intrinsic complexity of pricing and learning in the presence of competition.
\vspace{1cm}
\newpage
\section{Introduction}
\subsection{Motivation}
It is becoming increasingly common in today's online marketplaces that sellers' pricing decisions are determined by algorithms. The most striking example is arguably Amazon.com, which made more than 2.5 million price changes each day during 2013---a staggering figure that---most likely---has only increased ever since.\footnote{https://www.profitero.com/2013/12/profitero-reveals-that-amazon-com-makes-more-than-2-5-million-price-changes-every-day/, visited on December 12, 2017.} Even the price of the Bible---not the most obvious candidate for dynamic pricing---changes dozens of times each year,\footnote{https://camelcamelcamel.com/Holy-Bible-James-Version-Burgundy/product/0718015592, visited on December 12, 2017} which reveals that algorithmic pricing has gained a strong foothold in today's business practice. The complexities of optimally adjusting prices in response to competitors' prices, changing market circumstances, interactions between products in the seller's own portfolio, consumer reviews, incomplete information about consumers’ behavior, and many more factors that affect demand and revenue are obviously huge. To address these complexities, a large stream of scientific literature has emerged that designs pricing algorithms and analyzes their performance. A particularly large research area has evolved around the question of \emph{learning}: how should a seller price its products to optimize profit when the price-demand relation is unknown upfront, and therefore has to be learned from accumulating sales data? 

In recent years, a large number of studies have appeared that address this question from a monopolist's perspective (see literature review below). These research efforts have led to an understanding of the structure of optimal pricing strategies in a monopolist setting, in particular into the question of how much effort a seller should put into price experiments in order to strike the right balance between `exploration' (conducting price experiments in order to learn the price-demand relation) and `exploitation' (utilizing statistical knowledge to maximize profit). 

For pricing and learning in a competitive environment, the picture is rather different. It turns out to be very difficult to give a useful qualitative assessment of a pricing strategy, for the simple reason that its performance depends on the (unknown) pricing behavior of competitors. A particular strategy may work very well when used against simplistic pricing rules, but perform much worse against sophisticated algorithms. Even the right performance measure is not clear (can one, e.g., improve upon the full information Nash equilibrium?). Not only is theoretical understanding limited; there also does not appear to be an extensive numerical study that compares the practical performance of different algorithms.

Thus, there is a serious lack of understanding of the structure of well-performing pricing strategies with learning and competition, while at the same time understanding these pricing strategies is increasingly important from a practical viewpoint. 
This motivated the organizers of the INFORMS Revenue Management \& Pricing Section Conference 2017 to organize a dynamic pricing contest, in order to get insights into the numerical performance of different pricing strategies in a competitive environment with incomplete information, and so to gain insight into the properties of well-performing pricing policies. The results of this contest are reported in this paper.

\subsection{Literature}
The literature on `learning and earning' from a monopolist's perspective has gained much attention in recent years: see, e.g.,  \cite{AramanCaldentey2009,BesbesZeevi2009,FariasvanRoy2010,HarrisonKeskinZeevi2012,BroderRusmevichientong2012,
ChenFarias2013,SimchiLeviWangWeinstein2013, denBoerZwart2014a,KeskinZeevi2014,denBoerZwart2015}, and \cite{JohnsonFerreiraSimchiLeviWang2016}. A recent review of these and related references is provided by \cite{denBoer2015}.
A main take-away from this strand of literature is the importance of having the  `right' amount of \emph{price experimentation}.

The importance of incorporating competition into these learning-and-earning models, and the potential detrimental effect of ignoring competition, has been demonstrated by  \cite{SchinkelTuinstraVermolen2002, Tuinstra2004, BischiEtAl2004, BischiEtAl2007,IslerImhof2008, CooperHomemdeMelloKleywegt2009}, and \cite{AnufrievEtAl2013}, building forth on earlier work by Kirman \citep{Kirman1975,Kirman1983, Kirman1995, BrousseauKirman1992}. 

Various approaches have been adopted to incorporate competition into learning-and-earning problems. \cite{BertsimasPerakis2006} consider least-squares learning in an oligopoly with finite inventories and linear demand function, and propose an algorithm for estimation and pricing. \cite{KwonEtAl2009, LiYaoGao2011, ChungEtAl2012} adopt the framework of differential variational inequalities to study a capacitated oligopoly, propose an algorithm to solve these equations, and estimate unknown parameters via Kalman filtering. \cite{PerakisSood2006} (see also \cite{FrieszEtAl2012}) take a robust-optimization approach to the dynamic oligopoly pricing problem, and study Nash equilibrium policies. \cite{FisherEtAl2017} conduct a field experiment with randomized prices to estimate a consumer-choice model that does not require competitor sales observations, design a best-response pricing strategy, and test it with a field experiment for a leading Chinese online retailer. 

A sample from the extensive economics and econometrics literature that study asymptotic behavior of pricing strategies in competitive environments is \cite{CyertDeGroot1970, Kirman1975, AghionEspinosaJullien1993, MirmanSamuelsonUrbano1993, FishmanGandal1994, Harrington1995, BergemannValimaki1996, Gallego1998, DoloresAlepuz99, RassentiEtAl2000, BelleflammeBloch2001, KellerRady2003, DimitrovaSchlee2003}. These papers typically assume that a particular learning method is used by the competitors, and study whether the price process converges to a Nash equilibrium. 

The computer science literature also proposes several pricing-and-learning algorithms, see e.g.\ \cite{GreenwaldKephart1999, DasguptaDas2000, TesauroKephart2002, KutschinskiEtAl2003, Kononen2006, JumadinovaDasgupta2008,JumadinovaDasgupta2010,RamezaniBosmanLaPoutre2011}. For a further discussion of these and other relevant papers, we refer to Section 6.2 of \cite{denBoer2015}.

Finally, several simulation platforms have been designed to assess the performance of pricing policies, see, e.g., \cite{DiMiccoEtAl2003} or  \cite{Boissier2017}.

\subsection{Contributions}
This paper presents the results of the Dynamic Pricing Challenge, held on the occasion of the 17\textsuperscript{th} INFORMS Revenue Management and Pricing Section Conference on June 29-30, 2017, at the Centrum Wiskunde \& Informatica, Amsterdam, The Netherlands. For this challenge, participants were invited to submit pricing and learning algorithms that would compete for revenue in a broad range of simulated market environments in both duopoly and oligopoly settings. The extensive simulations that we ran allow us to describe the numerical performance of various pricing and learning algorithms and provides insight into the performance and properties of several types of policies. Given that the participants submitted a wide variety of algorithms---such as bandit-type models, customer choice models, econometric regression models, machine learning models, and greedy ad-hoc approaches---we are able to relate the performance of a broad range of algorithms to different market structures.

Hence, this paper offers a framework to analyze various paradigms from the field of pricing and learning with competition and allows us to consider market dynamics that are analytically intractable and can not be empirically analyzed due to practical complications. As such, this paper presents the results of a controlled field experiment that improve our understanding of pricing and learning with competition and helps to guide future research. Our most important findings are:
\begin{itemize}
\item Relative performance varies substantially across different market dynamics. Some algorithms perform well in competitive environments, whereas others are better at exploiting monopolist-like environments. None of the considered algorithms is able to dominate the others in all settings.
\item Relative performance varies substantially across oligopoly and duopoly markets. For example, algorithms based on linear demand models perform very well in duopoly competitions, whilst performing poorly in oligopolies. 
\item The algorithms that generate most revenue are very reliant on price-sensitive customers, making them vulnerable to intensified competition. Other algorithms are more robust in a sense that they were able to generate revenue from various types of customers and attract more loyal customers.
\item A greedy algorithm that follows the lowest-priced competitor in a tit-for-tat fashion proves very difficult to outperform.
\item Ignoring competition is increasingly harmful when competition is more fierce, i.e., when the number of competitors in the market is large and/or price sensitivity of the customers is high. 
\item The amount of exploration needs careful consideration as too much exploration hurts performance significantly.
\end{itemize}
The organization of the rest of this paper is as follows. In Section \ref{sec:expdesign} we describe the experimental design of this study. In Section \ref{sec:results} and \ref{sec:discussion} the results are presented and discussed, respectively. Finally, in Section \ref{sec:conclusion} some concluding remarks are provided. 
%
%
%
%
\section{Experimental Design}\label{sec:expdesign}
\subsection{Experimental Setting}\label{sec:exp_setting}
This experiment was designed to resemble a market in which the competitors all sell a single product to a group of heterogeneous customers. The competitors have no information a priori on either the demand mechanism or the behavior of the other competitors and are required to post a price before each (discrete) time period. Furthermore, it was assumed that competitors can monitor each other's prices, but only observe their own sales (we refer to sales as the number of items sold). This is true for many markets in reality, especially in online retailing, where retailers can monitor competitors' prices without much effort. Thus, the participants of this experiment were required to design an algorithm that would accept as input their own realized sales and the historical prices of all competitors and, subsequently, as output returns their price for the period to come. In addition, we assumed there are no inventory restrictions and, to give the participants some direction, the following domain knowledge was made available: ``it seems unlikely that posting prices higher than 100 is optimal''. For convenience and to prevent compatibility issues, all participants were required to submit their pricing policy in Python 2.7.x or 3.x and no restrictions on the use of libraries were put in place. 

To evaluate the performance of all submitted policies we ran 5000 simulations, where a single simulation consists of two different settings of competitive market environments:
\begin{itemize}
\item Duopoly competition: all competitors compete in a round-robin setup, i.e., each competitor competes with all other competitors in one-vs-one contests.
\item Oligopoly competition: all competitors compete simultaneously against each other.
\end{itemize}
Thus, if $m$ is the number of participating competitors, each simulation consists of $\binom{m}{2}$ duopoly competitions and one oligopoly competition (in this experiment there were eight competitors, thus $m=8$ and each simulation consists of $\binom{8}{2}=28$ duopolies and one oligopoly).
%
%
The oligopoly competition is especially insightful as it allows us to evaluate the competitors in a very competitive environment, whereas the duopoly competition can help us understand what the relative strength of the different competitors is. 

\begin{figure}
  \centering
  \includegraphics[width=1.0\textwidth]{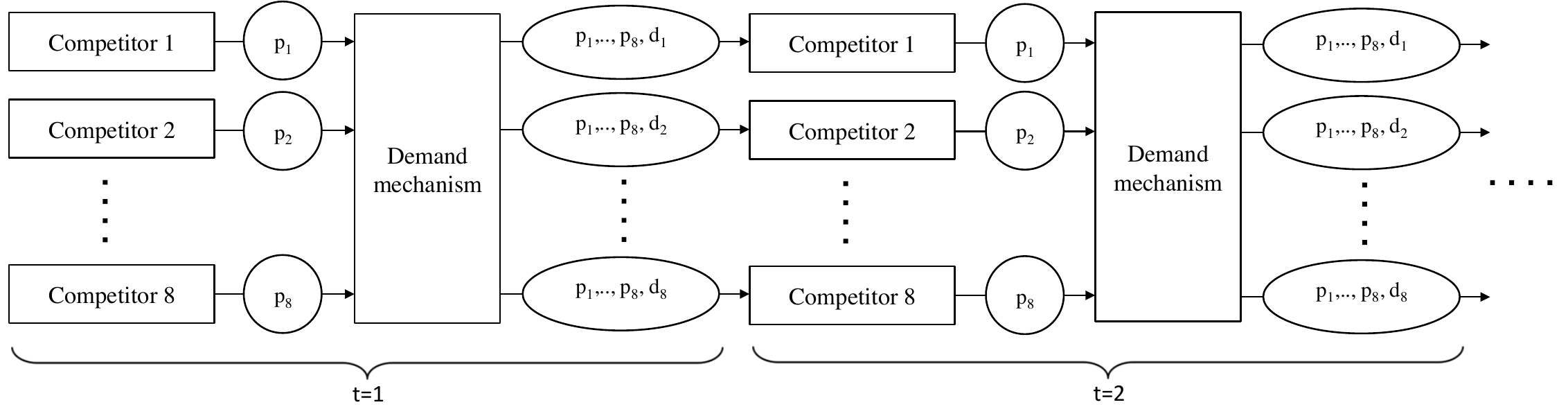}
  \caption{Overview of one simulation (oligopoly competition only); $p_i$ and $d_i$ denote the price and sales quantity of competitor $i$, respectively.}
  \label{fig:schema}
\end{figure}

In a single simulation, each of the duopoly competitions and the oligopoly competition consists of $1000$ discrete time periods. This means that for $t\in\{1,\hdots,1000\}$, each competitor posts a price for $t+1$ and, subsequently, sales quantities for period $t+1$ are generated from the undisclosed demand mechanism and all competitors earn revenue accordingly. This iterative process for the oligopoly competition is illustrated in Figure \ref{fig:schema}, whereas for the duopoly competitions, the scheme is the same except for the number of competitors, which is then two. 

Finally, it was communicated to all participants that we would construct a ranking and determine the winner as follows. In each simulation, for each competitor we compute its revenue share of that particular simulation by averaging
\begin{itemize}
\item the competitor's share of total revenue earned in the oligopoly competition and
\item the competitor's share of total revenue earned in the duopoly competitions.
\end{itemize}
The latter is computed by aggregating all the revenue earned in the $\binom{m}{2}$ duopoly competitions and computing the competitors' revenue shares accordingly. More precisely, if $x_{ij}$ is the revenue earned by competitor $j$ in the oligopoly competition of simulation $i$ and $y_{ijk}$ is the revenue earned by competitor $j$ in the duopoly competition versus competitor $k$ in simulation $i$, then competitor $j$'s revenue share of simulation $i$ is equal to
\begin{gather}\label{eq:revshare}
\frac{1}{2}(\bar{x}_{ij} + \bar{y}_{ij}), \text{ where }  \bar{x}_{ij} :=\frac{x_{ij}}{\sum_{k=1}^m x_{ik}} \text{ and } \bar{y}_{ij}:=\frac{\sum_{k=1}^m y_{ijk}}{\sum_{u=1}^m \sum_{k=1}^m y_{iuk}}.
\end{gather}
Thus $\bar{x}_{ij}$ and $\bar{y}_{ij}$ are the oligopoly and duopoly revenue shares, respectively, of competitor $j$ in simulation $i$. Note that, as such, a competitor's duopoly revenue share also depends on the revenue earned in competitions between other competitors. Consequently, it is not beneficial to earn a high revenue share in a duopoly competition where relatively little revenue is earned. The final score for competitor $j$ is simply the average over all its revenue shares, i.e., its final score equals $\frac{1}{5000}\sum_{i=1}^{5000} \frac{1}{2} (\bar{x}_{ij} + \bar{y}_{ij})$.

This way of constructing a final score is a design choice of the pricing contest; clearly, numerous alternative ways to measure performance are conceivable. 
%
\subsection{Competitor Algorithms}
In Table \ref{tab:algdescr} summaries of all the competing algorithms are provided (see appendix for more detailed descriptions). The competitors rely on a wide range of demand models:
\begin{itemize}
\item linear models: \ols and \wls (ordinary and weighted least squares, respectively),
\item bandit models: \mabgrid, \mabbuckets, and \mabmodels (bandits where the arms pertain to price points in a grid, price buckets, and demand models, respectively),
\item customer choice modeling: \logit,
\item machine learning: \ml,
\item greedy price-matching heuristic: \greedy.
\end{itemize}
All competitors randomize prices in the first periods and all competitors except \logit and \greedy also engage in exploration later on to capture possible non-stationary dynamics (which was not the case for the markets that we simulated). Regarding the modeling of competitor behavior, most competitors use variants of exponential smoothing to predict competitor prices and optimize own prices accordingly. Others model competitors' prices as multivariate normal random variables, use the median of historical prices as predictor, or ignore competition altogether. All non-bandit models use a line search to optimize their own revenue with respect to price, except for \wls, which optimizes own revenue relative to the revenue earned by the competition. 
\begin{landscape}
\begin{table}
\centering
\footnotesize
\begin{tabularx}{\linewidth}{ABCD}\toprule
Name & Demand Model & Pricing & Competitors' prices \\ \midrule
\textsc{logit} {\scriptsize[e,1,226]} & Finite mixture logit  &  During the first 100 periods the price is set as the minimum of the prices observed in the previous period. Then prices are optimized w.r.t.\ the demand model through a line search. & Multivariate normal distribution on historical prices \\ \midrule
\textsc{ols} {\scriptsize[f,2,114]}& Four linear regression models taking all combinations of price/log(price) and demand/log(demand). Select the model with the highest $R^2$. & During the first 40 periods and subsequently each time period with $5\%$ probability a price is uniformly sampled from the interval $(0,100)$. With probability $1\%$ price is set to 0. Otherwise, prices are optimized through a line search. &Ignored \\ \midrule
\textsc{b-grid} {\scriptsize[c,1,81]}& $\varepsilon$-greedy multi-armed bandit algorithm with ten arms pertaining to the prices $10,20,\hdots,100$. & With probability $\varepsilon=0.2$ an arm is chosen randomly, otherwise the arm with the highest average revenue so far is selected. & Ignored \\ \midrule
\textsc{b-bucket} {\scriptsize[c,2,160]}&$\varepsilon$-greedy multi-armed bandit algorithm with ten arms pertaining to the price buckets $(0,10],(20,30],\hdots,(90,100]$.&With probability 0.2 an arm is chosen randomly, otherwise the arm with the highest average revenue for the predicted competitor price (duopoly) or the mode of the competitors' prices (oligopoly) is chosen. & Exponential smoothing \\ \midrule
\textsc{b-model} {\scriptsize[c,2,1010]}&$\varepsilon$-greedy multi-armed bandit algorithm with four arms pertaining to three different demand models, each capturing different consumer behavior, and an epsilon-greedy algorithm in the fourth arm. & During the first 100 periods prices are set randomly, thereafter with probability 0.2 an arm is chosen randomly; otherwise offer the price suggested by the arm expected to generate the highest revenue, given a prediction of the competitors' prices. & Exponential smoothing \\ \midrule
\textsc{ml} {\scriptsize[d,1,325]} & Least-squares, ridge, Lasso, Bayesian ridge, and stochastic gradient descent regression, and random forest (model selection through cross-validation). & Exploration cycles are initiated regularly and consist of 40 time periods in which prices are set according to a cosine function around the mean price level observed. Otherwise prices are optimized w.r.t.\ the demand model through a line search. & Exponential smoothing \\ \midrule
\textsc{greedy} {\scriptsize[d,1,11]}& None & Price is set as minimum price observed in the previous period. If this price is lower than the 10\% percentile of all the prices observed in the last 30 time periods, then the price is set as the maximum of this percentile and 5. & Ignored \\ \midrule
\textsc{wls} {\scriptsize [g,1,387]}& Weighted least squares where weights capture time-dependent aspects of demand. & Prices are set to maximize own revenue relative to competition. Prices are randomized during the first 10 periods prices and when own prices are constant for three subsequent periods.  & Median of historical prices over variable window. \\ 
\bottomrule
\end{tabularx}
\caption{The variables in brackets after each competitor name corresponds to affiliation (see title page), team size, and lines of code, resp.}
\label{tab:algdescr}
\end{table}
\end{landscape}
\subsection{Demand Mechanism}\label{subsec:demandmechanism}
The design of the demand mechanism that we propose is built on the belief that it should resemble a competitive market with a heterogeneous customer base, as is often observed in practice. Meanwhile, we had to carefully manage the complexity of the demand mechanism to allow for evaluation, interpretation, and further analysis following the contest. In doing so, we assumed the arrival process and demand mechanism to be time-independent within a simulation, i.e., during time steps $1,\hdots,1000$ the arrival process is homogeneous and the demand function is static. Furthermore, we assumed that the market consists of three segments, namely loyal customers, shoppers, and scientists, who all have their own parameterized demand functions, as we will explain in the following sections. We emphasize once more that the participants of this competition were not aware of any of the aforementioned assumptions regarding the market structure. All in all, we realize that the outcomes inevitably depend on the ground truth that is constructed here, but it is intended to be versatile enough to reward the competitors that are best able to learn various types of demand dynamics.
\subsubsection{Arrivals and population composition}
We assume Poisson customer arrivals over time with mean arrivals per time period equal to  $\lambda_i$ in the $i^{\text{th}}$ simulation, where $\lambda_i\sim U(50,150)$. Here and throughout, if we write $x \sim F$, we mean that $x$ is sampled from (and not distributed as) $F$, i.e., $x$ is a realization. Furthermore, we denote the shares of the three customer segments, i.e., shoppers, loyals, and scientists, by $\theta_i^{\text{sho}}$, $\theta_i^{\text{loy}}$, and $\theta_i^{\text{sci}}$ in the $i^{\text{th}}$ simulation. In addition, the scientist segment is assumed to consist of two subsegments, namely PhDs and professors, with respective shares of $\gamma_i^{\text{phd}}$ and $\gamma_i^{\text{prof}}$. The sampling of arrivals is summarized in the code block titled `Arrival Process', 
\begin{algorithm}
\floatname{algorithm}{Arrival Process}
\caption{}
\label{alg:arrival}
\begin{algorithmic}
\FOR{$i \in \{1,\hdots,5000\}$}
\STATE Sample arrival rate $\lambda_i\sim U(50,150)$
\STATE Sample segment shares $\theta_i^{\text{sho}}$, $\theta_i^{\text{loy}}$, $\theta_i^{\text{sci}}$
\STATE Sample subsegment shares $\gamma_i^{\text{phd}}$, $\gamma_i^{\text{prof}}$
\FOR{$t \in \{1,\hdots,1000\}$}
\STATE Sample arrivals $n \sim \text{Poisson}(\lambda_i)$
\STATE Sample segment arrivals $n^{\text{sho}}, n^{\text{loy}}, n^{\text{sci}} \sim \text{Multinom}(n,(\theta_i^{\text{sho}}, \theta_i^{\text{loy}}, \theta_i^{\text{sci}}))$
\STATE Sample subsegment arrivals $n^{\text{phd}}, n^{\text{prof}} \sim \text{Multinom}(n^{\text{sci}},(\gamma_i^{\text{phd}}, \gamma_i^{\text{prof}}))$
\ENDFOR
\ENDFOR
\end{algorithmic}
\end{algorithm}
where $n$ is the number of arriving customers in a certain period, consisting of $n^{\text{sho}}$, $n^{\text{loy}}$, and $n^{\text{sci}}$ shoppers, loyal customers, and  scientists, respectively. The scientist segment, consists of $n^{\text{phd}}$ PhDs and $n^{\text{prof}}$ professors. Thus, the number of arriving customers $n$ equals $n^{\text{sho}} + n^{\text{loy}} + n^{\text{sci}}$ and the number of arriving scientists $n^{\text{sci}}$ equals $n^{\text{phd}} + n^{\text{prof}}$.
\subsubsection{Demand of loyal customers and shoppers}
The WTP of shoppers is assumed to be exponentially distributed with mean $\beta_i^{\text{sho}}$ in simulation $i$, where $\beta_i^{\text{sho}} \sim U(5,15)$. In each time period in simulation $i$, we sample a WTP for each arriving shopper from the exponential distribution with mean $\beta_i^{\text{sho}}$ and compare these WTPs to the lowest price offered in the market. Each shopper for whose WTP exceeds the lowest price offered, buys from the competitor that offers the lowest price and otherwise leaves without buying anything. Ties are broken randomly.

For the loyal customers, we assume their WTP is exponentially distributed as well, but with mean $\beta_i^{\text{loy}}$ in simulation $i$, where $\beta_i^{\text{loy}}=u \cdot \beta_i^{\text{sho}}$ and ${u\sim U(1.5, 2.0)}$, making the loyal customers relatively price-insensitive compared to the other customer groups, as one would expect from a loyal customer segment. In the simulation, each loyal customer is assigned randomly to a competitor, and they will buy from this competitor if their WTP exceeds the price the competitor is offering; otherwise they leave without making a purchase. Subsequently, we sample for all loyal customers a WTP and compare this to the price offered by the competitor that they are loyal to. In case their WTP exceeds the price offered, they buy, otherwise they leave without buying anything. 
\subsubsection{Demand scientists}
The demand of the scientists is assumed to follow a finite mixture logit model, or latent class logit model, where the mixture comprises professors and PhDs. Thus, the professors and PhDs both choose according to a logit model with their own parameters. Let the posted prices in period $t$ be equal to $p_t$ and define the probability that an arriving PhD purchases from competitor $k$ during simulation $i$ as follows
\[ q^{\text{phd}}_k(p_t) = \frac{ \exp\left({\alpha^{\text{phd}}_{i}} - \beta^{\text{phd}}_{n,i} \cdot p_{k,t}\right)}{1 + \sum_{j=1}^n \exp\left({\alpha^{\text{phd}}_{i}} - \beta^{\text{phd}}_{n,i} \cdot p_{j,t}\right) } \]
where $n=2$ in the duopoly case and $n=m(=8)$ in the oligopoly case. This probability is defined similarly for professors. The parameters $\alpha^{\text{phd}}_i$, $\beta^{\text{phd}}_{n,i}$, $\alpha^{\text{prof}}_i$, and $\beta^{\text{prof}}_{n,i}$ of simulation $i$ are sampled so that the prices that would be optimal for the shoppers, loyals, PhDs, and professors in isolation are of the same order of magnitude. This is done to ensure that no unrealistically large differences in optimal prices between (sub)segments occur. In doing so, for the PhDs we set $\alpha^{\text{phd}}_{i}$ equal to $\beta_i^{\text{sho}}$ and, subsequently, set $\beta^{\text{phd}}_{n,i}$ such that the optimal price for the PhDs is within 50\% of the optimal price for the shoppers (which is equal to $\beta_i^{\text{sho}}$). This is achieved as follows
\begin{gather*}
\alpha^{\text{phd}}_{i} = \beta_i^{\text{sho}} \\
p^{\text{phd}}_i := \beta_i^{\text{sho}} \cdot u,\ \text{where } u \sim U(0.50,1.50) \\ 
\beta^{\text{phd}}_{n,i} = \frac{W\left(n e^{\alpha^{\text{phd}}_{i}-1}\right) + 1}{p^{\text{phd}}_i}
\end{gather*} 
where $W$ is the Lambert function, i.e., $W(xe^x) = x$, which is uniquely defined in this case as $n e^{\alpha^{\text{phd}}_{i}-1}> 0$. Thus, we set the parameters so that if the market consisted solely of PhDs, then, overall revenue would be maximized if $p^{\text{phd}}_i$ was set by all competitors, where $p^{\text{phd}}_i$ is within a reasonable distance of the average WTP of shoppers. Thus, the parameters are set so that 
\[ \argmax_{p\in\mathbb{R}^n_+} \sum_k p_k q^{\text{phd}}_k (p) = \mathbbm{1}_n p^{\text{phd}}_i \]
where $\mathbbm{1}_n p^{\text{phd}}$ is understood to be an $n$-vector with each element equal to $p^{\text{phd}}_i$.  Similarly, we set 
\begin{gather*}
\alpha^{\text{prof}}_{i} = \alpha^{\text{phd}}_{i}  \cdot u,\ \text{where } u\sim U(1.00,1.25) \\ 
p^{\text{prof}}_i := p^{\text{phd}}_i  \cdot u,\ \text{where } u\sim U(1.00,1.50) \\ 
\beta^{\text{prof}}_{n,i} = \frac{W\left(n e^{\alpha^{\text{prof}}_{i}-1}\right) + 1}{p^{\text{prof}}_i}
\end{gather*} 
so that the optimal price in a market consisting of only professors would be higher than in a market that consists solely of PhDs, while keeping the price levels in line. 
\subsubsection{Demand mechanism summary}\label{sec:mech_summary}
Summarizing, in simulation $i\in\{1,\hdots,5000\}$, in each period we see in expectation $\lambda_i\cdot\theta_i^{\text{sho}}$ arriving shoppers, $\lambda_i\cdot\theta_i^{\text{loy}}$ arriving loyals, $\lambda_i\cdot\theta_i^{\text{sci}}\cdot\gamma_i^{\text{phd}}$ arriving PhDs, and $\lambda_i\cdot\theta_i^{\text{sci}}\cdot\gamma_i^{\text{prof}}$ arriving professors. Each of these customer types chooses according to its own parameterized demand function (of which the parameters are constant throughout the simulation) as described in the previous sections. This all implies that in some simulations, $\theta_i^{\text{loy}} \approx 1$ meaning that all competitors are essentially monopolists, which should theoretically lead to higher prices, lower sales, and relatively high revenues. Another extreme, when $\theta_i^{\text{sho}} \approx 1$, resembles a market in which there is perfect competition---each competitor offers the same product to arriving customers that purchase the cheapest alternative available. In this case, one would expect prices to spiral down over time. In addition, the share of loyal customers in the market ($\frac{1}{3}$ on average) is independent of the number of competitors in the market. Therefore, in a duopoly, an arriving customer is loyal to a specific competitor with probability $\frac{1}{2} \cdot \frac{1}{3}$, whereas, in the oligopoly, this probability equals $\frac{1}{m} \cdot \frac{1}{3}$. Therefore, by construction, we anticipate the market to be much more competitive in the oligopoly setting than in the duopoly case, as one would expect from economic theory.
\section{Results}\label{sec:results}
First, in Section \ref{ssec:overall} a summary of the overall results is presented and then, in Sections \ref{ssec:oli} and \ref{ssec:duo}, the results of the oligopoly and duopoly competitions, respectively, are considered in greater detail. Thereafter, in Section \ref{sec:discussion}, a discussion is provided on the observations presented here. 
\subsection{Overall Results}\label{ssec:overall}
\begin{figure}
\centering
\includegraphics[width=\textwidth]{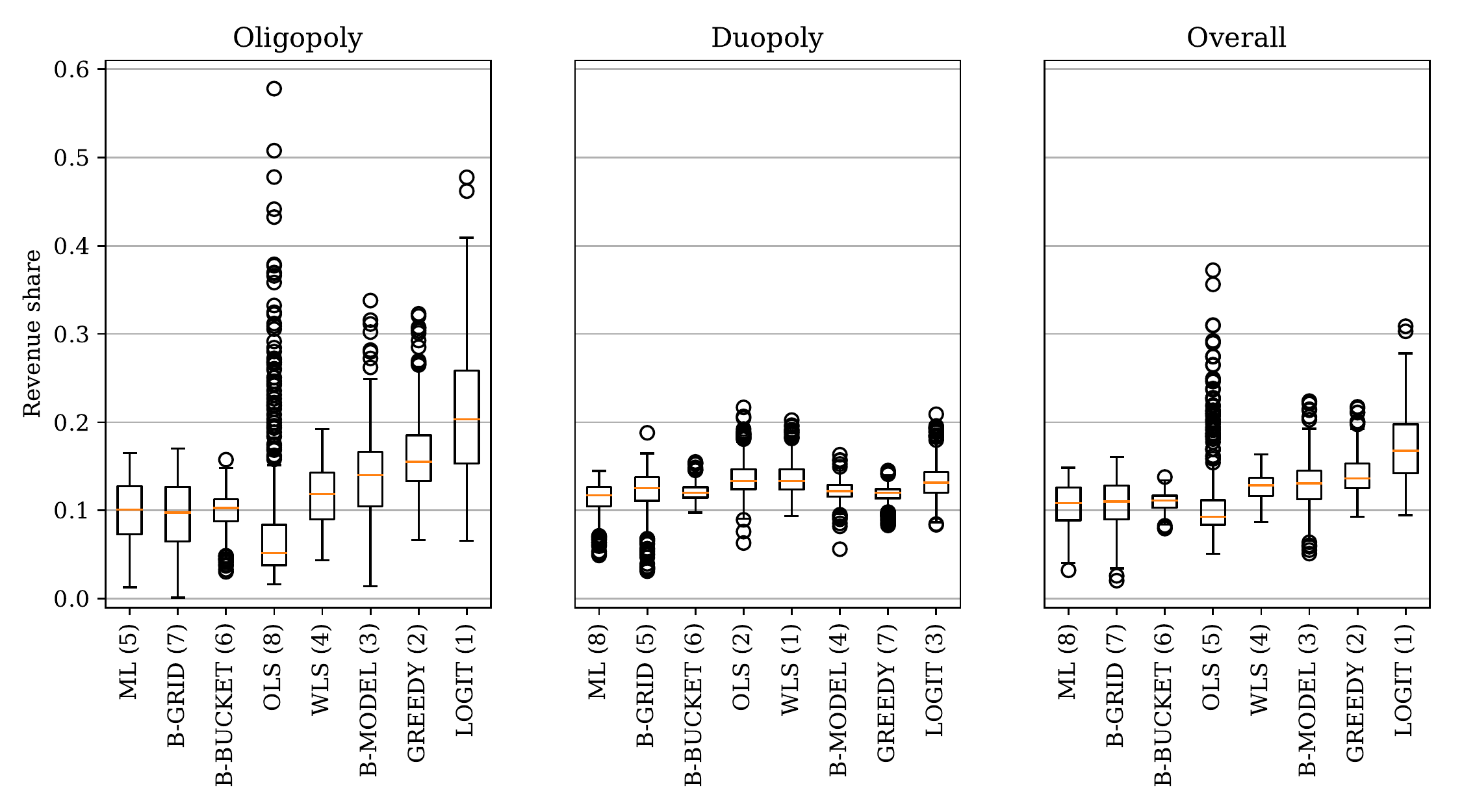}
\caption{Boxplots of the revenue shares for the oligopoly competition, the duopoly competitions, and the overall competition.
In parenthesis after the names of the algorithms are the rankings per competition part (e.g., \wls was the winner of the duopoly part). Only the first $500$ of $5000$ simulations are used to construct the boxplots for the sake of readability.
}
\label{fig:revbars}
\end{figure}

In Figure \ref{fig:revbars} boxplots of the revenue shares for the oligopoly, the duopoly, and the overall competition are provided. Thus, the boxplots in the left panel are based on the $\bar{x}_{ij}$'s, in the middle panel on the $\bar{y}_{ij}$'s, and in the right panel on $\frac{1}{2} (\bar{x}_{ij} + \bar{y}_{ij})$'s, all defined in \eqref{eq:revshare}. 
The figure reveals that \logit was the eventual winner of the competition and that its success was primarily due to its superior performance in the oligopoly competitions. Overall, the differences in performance in the oligopoly are substantial, whereas in the duopoly competitions it has proven to be much harder to outperform one another. Nevertheless, we observe remarkable differences in relative performance across the oligopoly and duopoly parts. For example, \ols nearly earned the highest mean revenue share in the duopoly competitions, while its performance in the oligopoly competition was on average the worst amongst all competitors. Note however, that despite this poor average, \ols did perform particularly well on some occasions (earning almost 60\% of all revenue on one occasion). The other way around, we observe that competitor \greedy performs poorly in the duopoly competitions, but that its performance in the oligopoly competition is relatively good, being second placed only after \logit. 
Furthermore, we observe that \mabbuckets, \mabgrid, and \ml are consistently outperformed in both duopoly and oligopoly competitions, as they are amongst the bottom four in both parts of the competition. In the following sections, we analyze the aforementioned observations in greater detail and make more detailed comparisons between the different pricing strategies in the market.
\subsection{Oligopoly Competition}\label{ssec:oli}
In this section, we uncover what causes the substantial differences in performance in the oligopoly competition, which are observable in the first panel in Figure \ref{fig:revbars}. This is done by analyzing how the competitors differ in terms of realized sales and prices posted and how their performance varies as the market composition, i.e., the shares of segments, differs across simulations. 

\begin{figure}
\centering
\includegraphics[width=\textwidth]{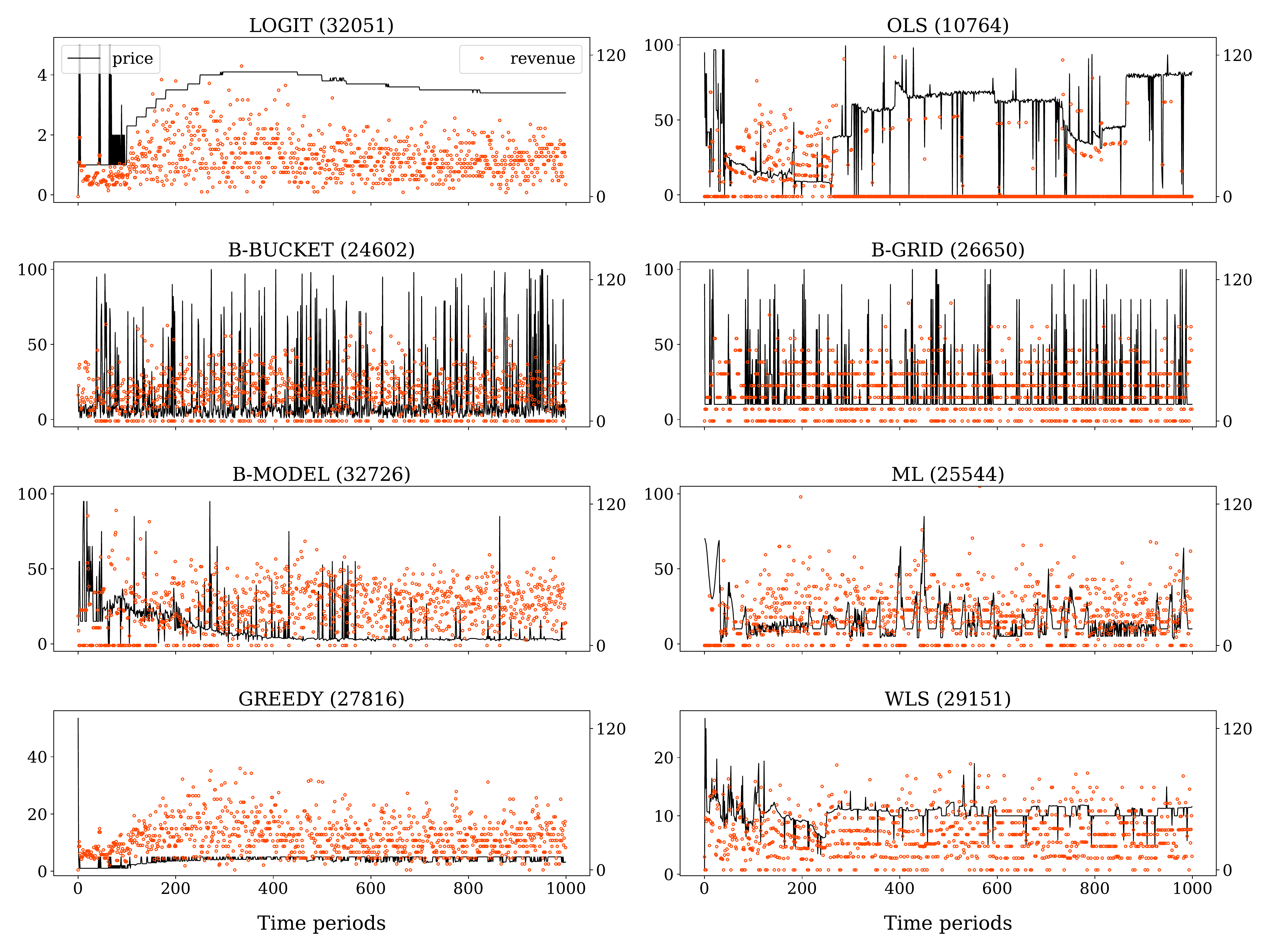}
\caption{Realizations of prices and revenue for simulation 4955. The solid line represents the prices and pertains to the left axis. The red dots represent the revenue earned and pertain to the right axis. In parenthesis is the total accumulated revenue. In this simulation the segment shares were equal to $\theta^{\text{sho}} =0.05$, $\theta^{\text{loy}} = 0.72$, and $\theta^{\text{sci}} = 0.23$.}
\label{fig:fc_realization}
\end{figure}
First of all, in Figure \ref{fig:fc_realization} we present the price and revenue realizations of a single simulation to demonstrate what a simulation typically looks like. Although a single simulation is not representative for performance in general, we found that the characteristics that can be observed in Figure \ref{fig:fc_realization} are demonstrative for most of the scenarios that we visually inspected. In particular, it can be observed that the prices of \logit, \mabmodels, \greedy, and \wls converge (after engaging in price exploration), whilst the other competitors show more erratic price paths. For example, \mabbuckets and \mabgrid put a lot of emphasis on exploration throughout the simulation without eventually converging to a small price range, which can be explained by the fact that these algorithms engage in active price exploration with a 20\% probability in each period. Furthermore, we see that \ols initially seems to converge to a competitive price, but that from around period 300 onwards it fluctuates around a relatively high price level, earning hardly any revenue. Finally, regarding \ml we can observe that it initiates many exploration cycles around a cosine function, which affects its performance negatively in this stationary environment.

\begin{figure}
\centering
\subfloat[]{\includegraphics[width=0.48\textwidth]{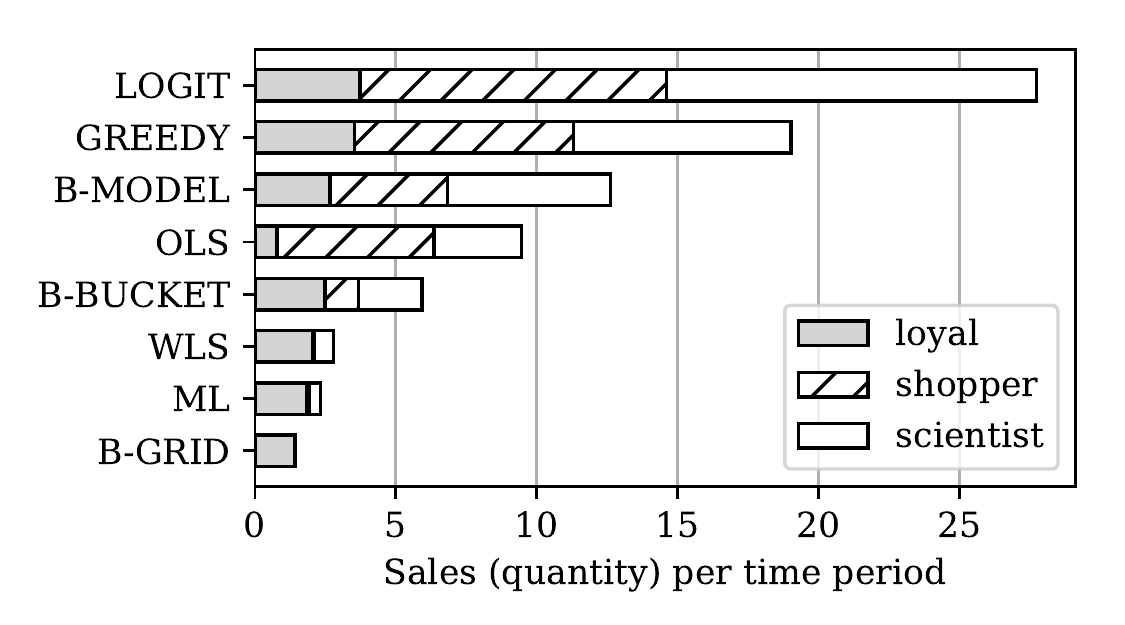}}
\subfloat[]
{\includegraphics[width=0.48\textwidth]{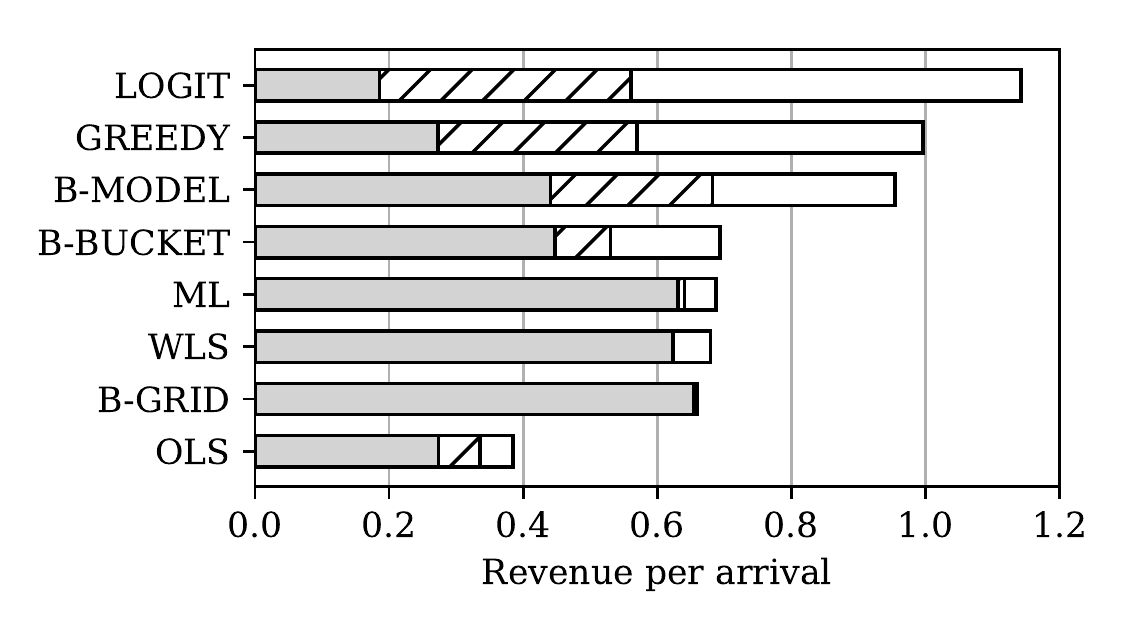}}
\caption{For the oligopoly competition (a) mean sales per time period and (b) mean revenue per arriving customer, split out over the three customer segments.}
\label{fig:sales_rev_per_arrival}
\end{figure}

To analyze how the competitors differ in sales generated across segments, in Figure \ref{fig:sales_rev_per_arrival} (a) the mean sales per time period per segment for each competitor is presented (note that sales is understood to be a quantity here and throughout). From this figure it can be observed that the three best-performing algorithms in the oligopoly, \logit, \greedy, and \mabmodels, are also able to generate the highest sales. Furthermore, \logit not only generates the highest total sales, but it is also able to generate the highest sales per customer type. This could well be due to the fact that if a competitor generates high sales in the shopper segment, this means that it is frequently the lowest priced competitor in the market, which, in turn, means that the scientists and its loyal customers are also likely to buy. Thus, high sales in the shopper segment leads automatically to relatively high sales in the other two segments. In addition, it is remarkable that \wls generates much lower sales than \ols, while, according to Figure~\ref{fig:revbars}, \wls performs significantly better in terms of revenue generation. Arguably, this is due to the fact that \wls sells predominantly to the high-paying loyal segment, whereas \ols generates its revenue primarily from the shopper and scientist segments, which are more price sensitive. Overall, the competitors that are able to reach the shoppers and scientists are capable of generating high sales, which concurs with the observation that for each competitor the potential sales from shoppers and scientists is much larger than from loyal customers (as was discussed in the final part of Section \ref{sec:mech_summary}).

\begin{figure}
\centering
\includegraphics[width=\textwidth]{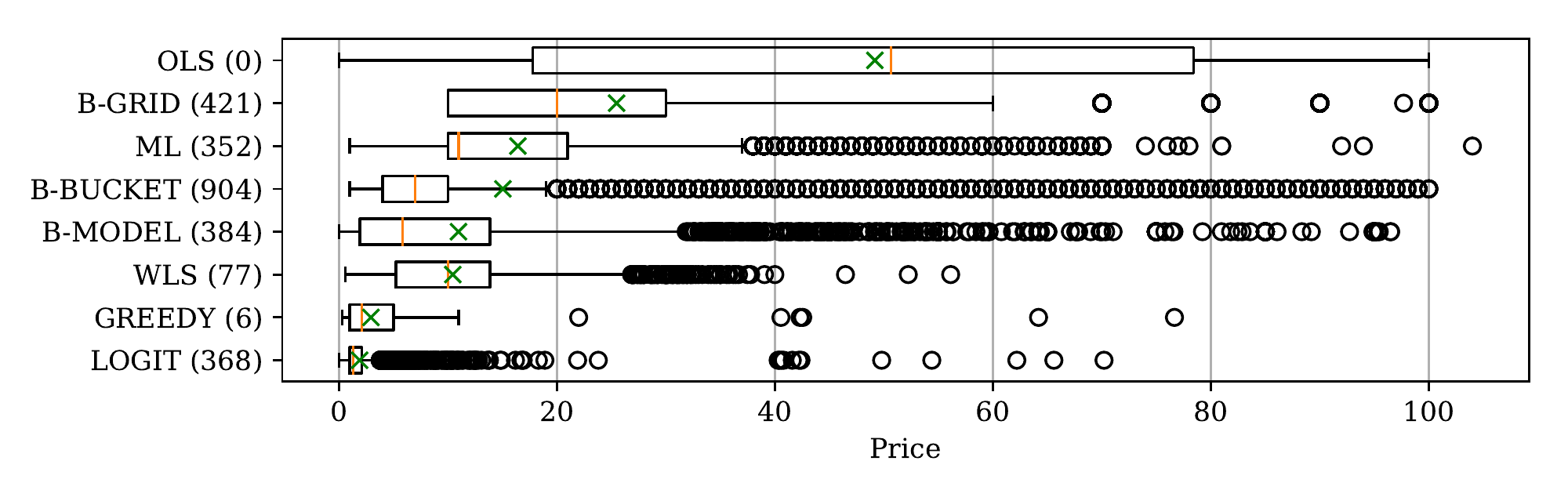}
\caption{Boxplot based on the prices posted in 5000 randomly chosen time periods (out of $5000 \times 1000$ time periods in total). The bars and crosses denote the median and mean, respectively. In brackets are the corresponding number of outliers.}
\label{fig:boxplot_prices}
\end{figure}

In Figure \ref{fig:sales_rev_per_arrival} (b) the mean revenue per arriving customer for each competitor is illustrated (e.g., \logit makes on average about 0.30 from an arriving shopper, whereas \mabgrid practically earns nothing in this case). The figure reveals that the competitors' distributions of earnings over the customer segments varies substantially. This can be seen when considering the top performers (\logit, \greedy, and \mabmodels), which make the majority of their income from shoppers and scientists. For the other algorithms, especially \ml, \wls, and \mabgrid, it holds that they earn hardly anything from shoppers and scientists, but are able to earn relatively much from incoming loyal customers. To relate these observations to the prices posted in the market, we provide in Figure \ref{fig:boxplot_prices} boxplots of the prices posted and in Table \ref{tab:extremes}, amongst other things, the fraction of time periods in which a competitor was the lowest or highest priced competitor in the market. In general, we observe that the successful algorithms set relatively low prices, earn most revenue from shoppers and scientists, and relatively little revenue from the loyal customers. It is by no means immediate that \logit's low-pricing strategy works well under all circumstances---its dependency on shoppers and scientists makes \logit vulnerable in case the market consists of more competitors that are aggressive on price. Thus, in a sense is \mabmodels more robust by being less reliant on the price sensitive segments. This is also illustrated by \mabgrid, which only serves loyal customers and generates the least sales, but does not earn substantially less than \wls, \ml, and, \mabbuckets and performs substantially better than \ols.
\begin{table}
\centering
\begin{tabular}{l|cc|cc|cc|} \toprule
&\multicolumn{2}{c|}{Price}&\multicolumn{2}{c|}{Sales}&\multicolumn{2}{c|}{Revenue}\\
&Low&High&Low&High&Low&High\\ \midrule
\logit&\underline{0.55}&0.00&0.01& \underline{0.37} &0.02&\underline{0.23}\\
\ols&0.13&\underline{0.60}&\underline{0.69}&0.13&\underline{0.68}&0.14\\
\mabbuckets&0.08&0.09&0.27&0.07&0.27&0.07\\
\mabgrid&0.00&0.19&0.47&0.00&0.44&0.10\\
\mabmodels&0.20&0.06&0.23&0.19&0.24&0.15\\
\ml&0.01&0.08&0.36&0.01&0.35&0.08\\
\greedy&0.46&0.00&0.06&0.27&0.07&0.15\\
\wls&0.00&0.01&0.17&0.00&0.16&0.09\\
\bottomrule
\end{tabular}
\caption{Fraction of time periods (of $5000 \times 1000$ time periods total) in which a competitor was lowest/highest priced, generated the lowest/highest sales, and earned the lowest/highest revenue. The column sums are greater than one because ties occur.}
\label{tab:extremes}
\end{table}

Regarding price experimentation, we observe from Figure \ref{fig:boxplot_prices} that \ols's policy induces a very wide price range, with its first quartile around twenty and the third quartile just below eighty, resembling a Gaussian distribution of prices. In addition, \greedy does not engage in much price experimentation, as one would expect from its construction. The other boxplots indicates that the price distributions all have a heavy right tail; these competitors engage in exploration coincidentally, while pricing around a relatively small interval for most of the time. 

\begin{figure}
\centering
\includegraphics[width=0.6\textwidth]{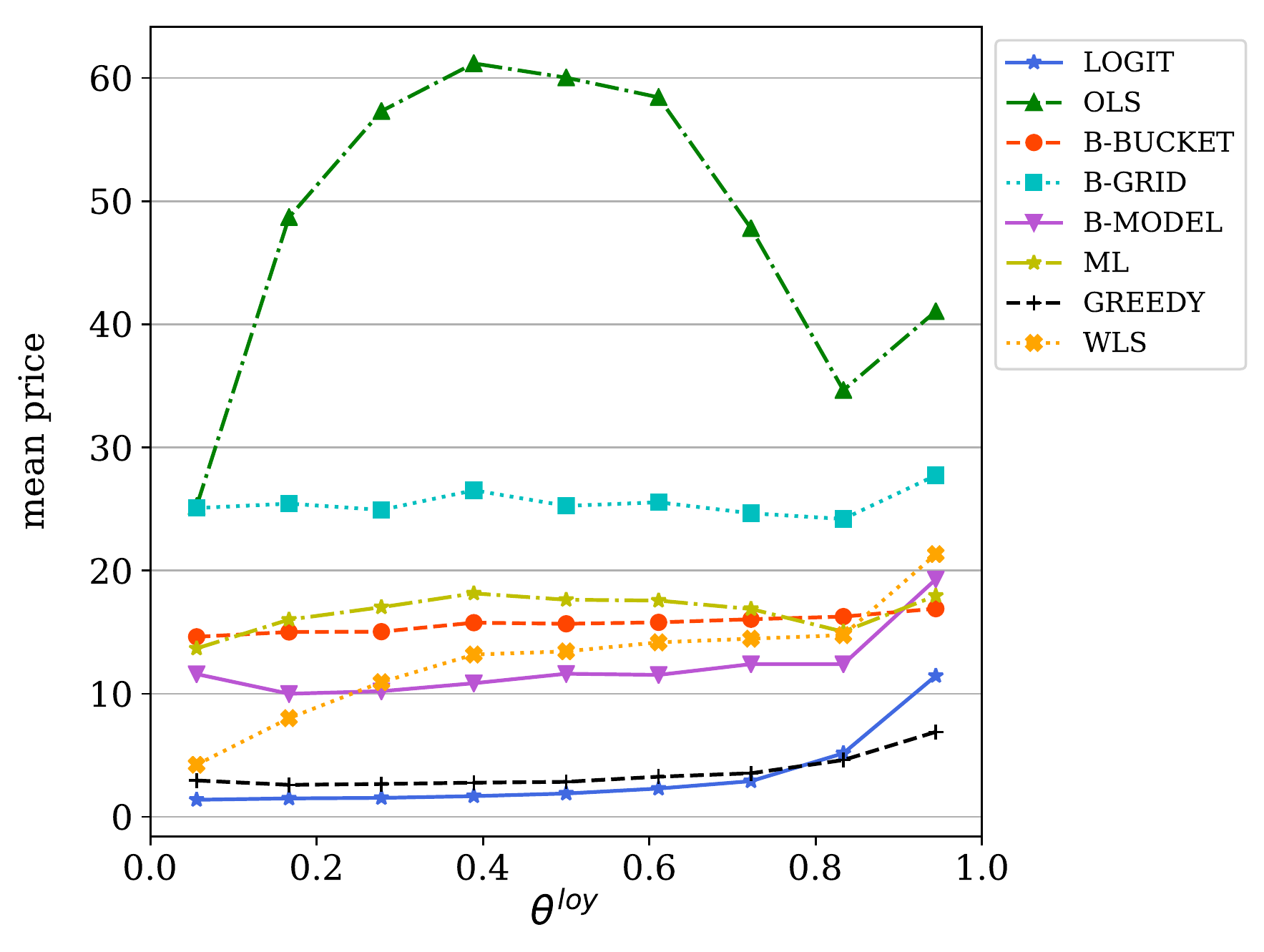}
\caption{Mean prices for various levels of the share of loyal customers $\theta^{\text{loy}}$.}
\label{fig:p_vs_loyalshareI}
\end{figure}

In Figure \ref{fig:p_vs_loyalshareI} the mean prices for different values of $\theta^{\text{loy}}$ are given.
This is insightful, since when the share of loyal customers  increases from zero to one, the market moves from a very competitive market to a market in which every competitor is essentially a monopolist. One would theoretically expect that competitors post higher prices once their pricing power increases. However, from the figures, we can observe that the prices of three of the four worst-performers, namely \ols, \mabgrid, and \ml, do not increase in the share of loyal customers. On the other hand, the prices of \logit, \greedy, \wls, \mabbuckets, and \mabmodels, do increase in the share of loyal customers. This indicates that these competitors are better capable of identifying the market structure and improve pricing decisions accordingly.

\begin{figure}
\begin{minipage}[t]{0.48\textwidth}
\includegraphics[width=\textwidth]{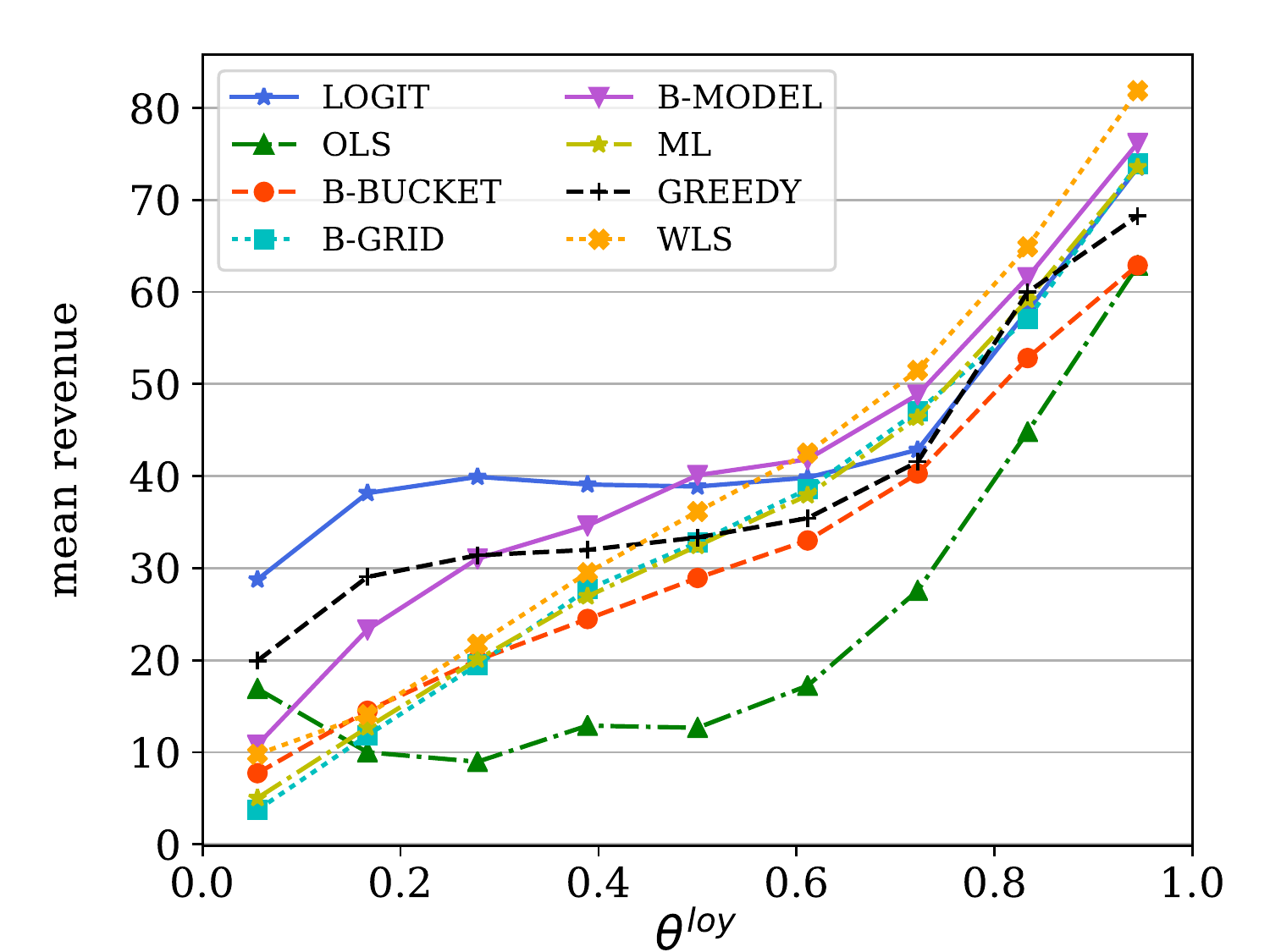}
\caption{Mean revenue per time period for various levels of $\theta^{\text{loy}}$.}
\label{fig:r_vs_loyalshare}
\end{minipage}
\hfill
\begin{minipage}[t]{0.48\textwidth}
\includegraphics[width=\textwidth]{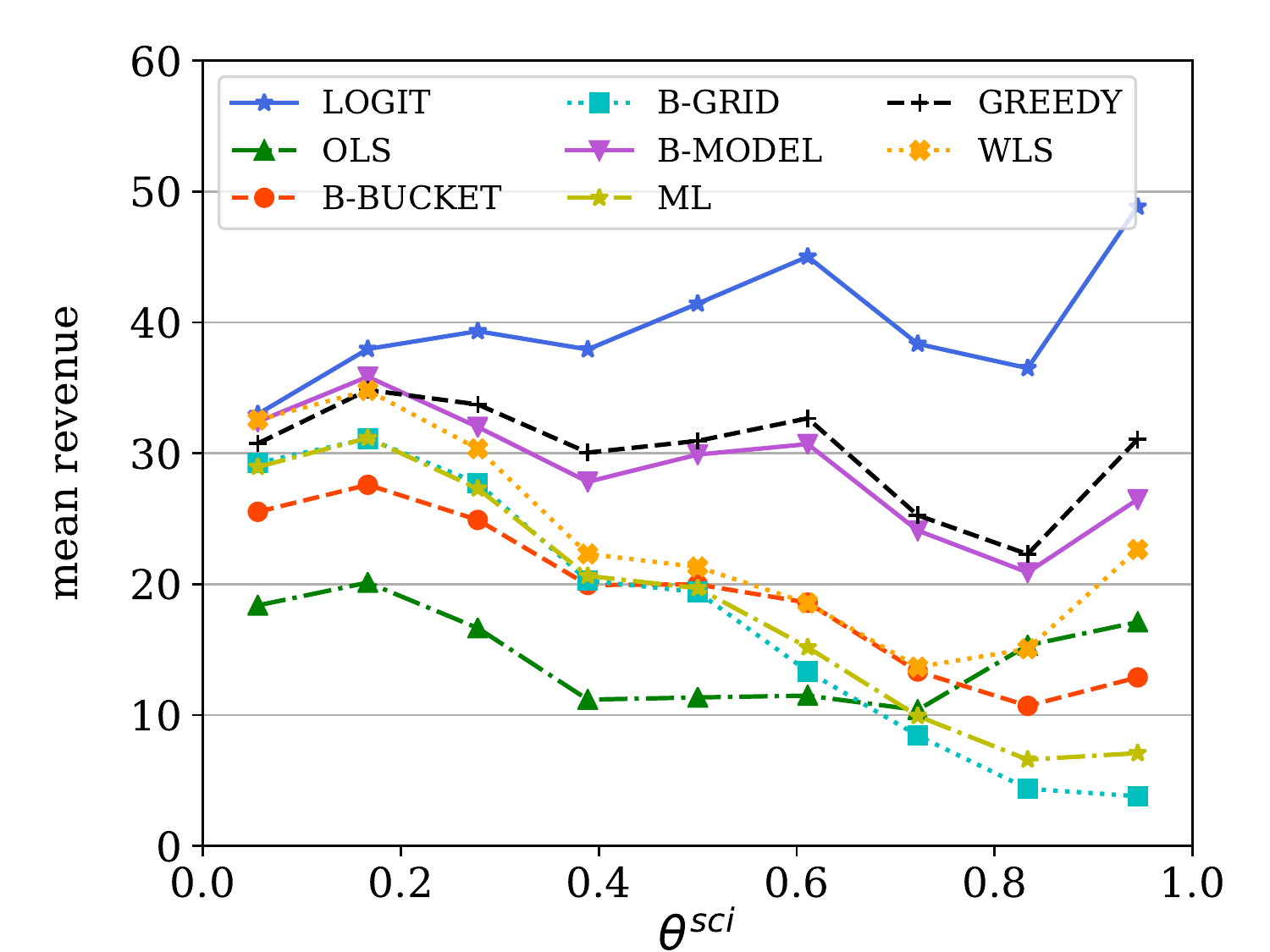}
\caption{Mean revenue per time period for various levels of $\theta^{\text{sci}}$.}
\label{fig:r_vs_scientistshare}
\end{minipage}
\end{figure}

Finally, Figures \ref{fig:r_vs_loyalshare} and  \ref{fig:r_vs_scientistshare} illustrate the mean revenue per time period for various segment shares of loyal customers, and scientists, respectively.\footnote{The figure for the shopper segment is very similar to Figure \ref{fig:r_vs_loyalshare} except for that the revenues decrease in the shopper share --- figure is omitted to save space.} From Figure \ref{fig:r_vs_loyalshare} we observe that, in general, revenues increase as the share of loyal customers increases, as one would theoretically expect. However, some competitors are better capable of exploiting the increase in pricing power than others---for example \wls's relative performance improves substantially as $\theta^{\text{loy}}$ increases, while \logit's relative performance deteriorates. On the other hand, in Figure \ref{fig:r_vs_scientistshare} we observe that performance across competitors diverges as the share of scientists increases. Most notably, \logit's relative performance increases substantially, which can be attributed to the fact that \logit's demand specification is able to closely resemble the demand function of the scientists (which is a finite mixture of logit demand functions).
%
\subsection{Duopoly Competition}\label{ssec:duo}
From Figure \ref{fig:revbars} it follows that in the duopoly competitions the differences in performance are less substantial than in the oligopoly competition. Two reasons can be identified why this is the case. First of all, the share of loyal customers is relatively large in the duopoly, so that both competitors have in general more pricing power whilst only having to consider one other competitor, making it harder to outperform one another. Second of all, the performance is not transitive in a sense that ``if A beats B and B beats C, then A beats C'', so that differences in performance tend to cancel out over the duopolies. Nonetheless, the duopolies are interesting to analyze the relative performance of the various pricing policies.
\begin{figure}
\centering
\includegraphics[width=\textwidth]{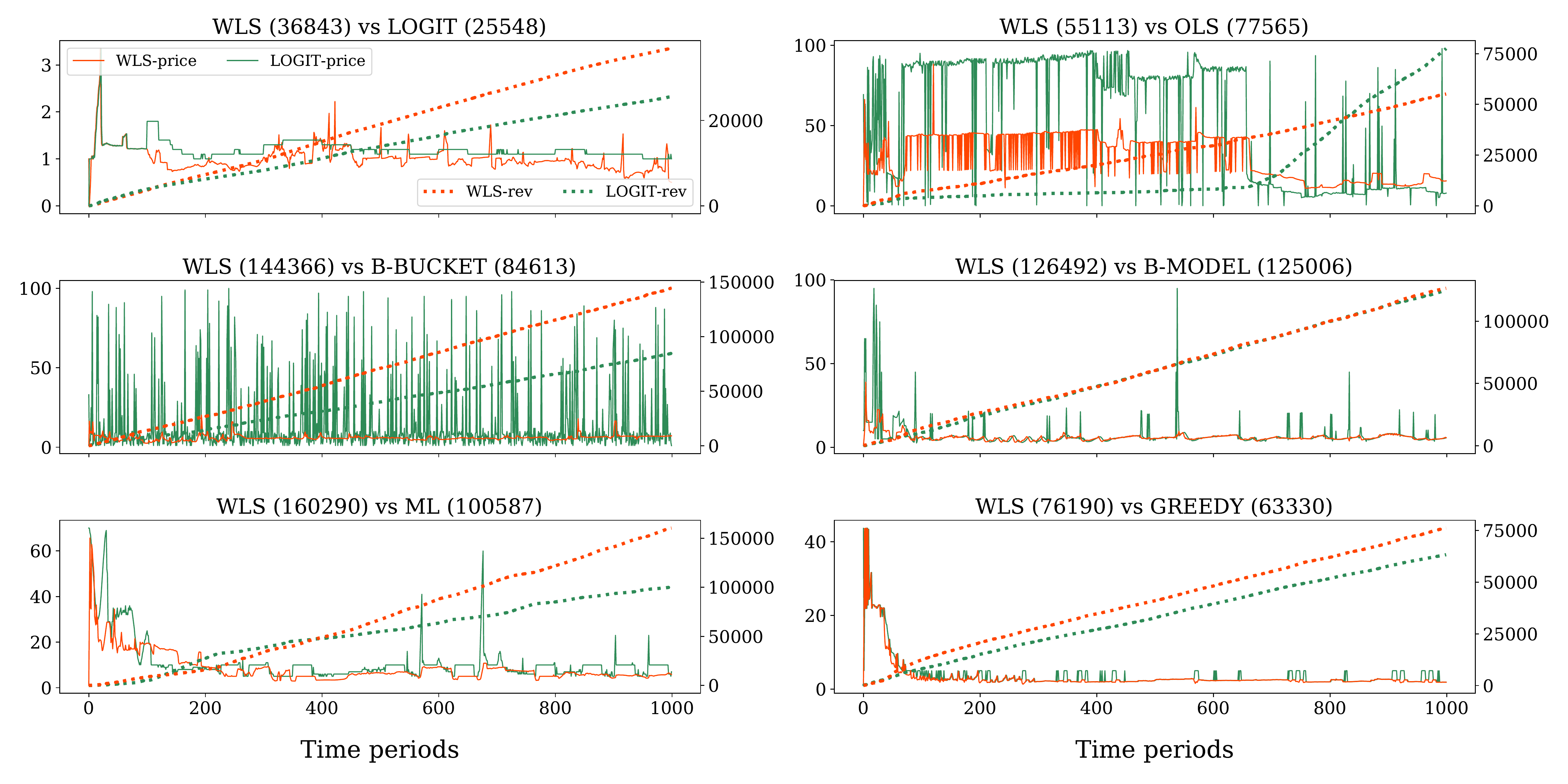}
\caption{Realizations of prices and revenue for simulation 2285. The red and green solid lines represent the prices of WLS and the corresponding opponent, respectively, and pertains to the left axis. The red and green dotted lines represent the cumulative revenue earned by WLS and the corresponding opponent, respectively, and pertains to the right axis. In parenthesis is the total accumulated revenue. In this simulation the segment shares were equal to $\theta^{\text{sho}} =0.33$, $\theta^{\text{loy}} = 0.29$, and $\theta^{\text{sci}} = 0.38$.}
\label{fig:rr_realization}
\end{figure}

For a single simulation, in Figure \ref{fig:rr_realization} we provide the realized prices and revenue of duopolies in which \wls, the top performer in the duopoly part, was involved.\footnote{A plot of \wls vs \mabgrid is omitted to save space but is very similar to the plot of \wls vs \mabbuckets.} The figure reveals that in the duopoly in which \wls and \logit compete relatively little revenue is generated since both algorithms price very low---this is not only the case in this example, but is a structural property, as will be shown below. It is interesting to see that \greedy's performance is weakened by its mechanism that resets prices to 5 if prices get too low (these are the bumps that are visible in the plot). Finally, as was the case in the oligopoly, \ols is in this case not able to find a competitive price level consistently (however it was still able to perform well in the duopolies).

In Table \ref{tab:duo_revenues} the mean revenue per time period for all the duopolies are provided.
\begin{table}
\centering
\small
\begin{tabular}{r|cccccccc|c} \toprule
&\rot{\logit}&\rot{\ols}&\rot{\mabbuckets}&\rot{\mabgrid}&\rot{\mabmodels}&\rot{\ml}&\rot{\greedy}&\rot{\wls}&\rot{average}\\ \midrule
\logit&&\cellcolor{YellowGreen}$252$&\cellcolor{YellowGreen}$221$&\cellcolor{YellowGreen}$319$&\cellcolor{Salmon}$238$&\cellcolor{Salmon}$242$&\cellcolor{Salmon}$266$&\cellcolor{Salmon}$96$&234\\
\ols&\cellcolor{Salmon}$235$&&\cellcolor{Salmon}$172$&\cellcolor{YellowGreen}$265$&\cellcolor{Salmon}$210$&\cellcolor{YellowGreen}$295$&\cellcolor{Salmon}$256$&\cellcolor{Salmon}$249$&240\\
\mabbuckets&\cellcolor{Salmon}$175$&\cellcolor{YellowGreen}$249$&&\cellcolor{YellowGreen}$256$&\cellcolor{YellowGreen}$207$&\cellcolor{YellowGreen}$272$&\cellcolor{Salmon}$198$&\cellcolor{Salmon}$175$&219\\
\mabgrid&\cellcolor{Salmon}$196$&\cellcolor{Salmon}$247$&\cellcolor{Salmon}$169$&&\cellcolor{Salmon}$239$&\cellcolor{YellowGreen}$306$&\cellcolor{YellowGreen}$274$&\cellcolor{Salmon}$181$&230\\
\mabmodels&\cellcolor{YellowGreen}$250$&\cellcolor{YellowGreen}$231$&\cellcolor{Salmon}$185$&\cellcolor{YellowGreen}$241$&&\cellcolor{YellowGreen}$270$&\cellcolor{Salmon}$176$&\cellcolor{Salmon}$218$&224\\
\ml&\cellcolor{YellowGreen}$249$&\cellcolor{Salmon}$221$&\cellcolor{Salmon}$177$&\cellcolor{Salmon}$236$&\cellcolor{Salmon}$203$&&\cellcolor{YellowGreen}$216$&\cellcolor{Salmon}$186$&213\\
\greedy&\cellcolor{YellowGreen}$267$&\cellcolor{YellowGreen}$260$&\cellcolor{YellowGreen}$206$&\cellcolor{Salmon}$273$&\cellcolor{YellowGreen}$196$&\cellcolor{Salmon}$209$&&\cellcolor{Salmon}$123$&219\\
\wls&\cellcolor{YellowGreen}$153$&\cellcolor{YellowGreen}$257$&\cellcolor{YellowGreen}$234$&\cellcolor{YellowGreen}$361$&\cellcolor{YellowGreen}$219$&\cellcolor{YellowGreen}$317$&\cellcolor{YellowGreen}$140$&&240\\\midrule
average&218&245&195&279&216&273&218&175&\\
\bottomrule
\end{tabular}
\caption{The pairwise mean revenue per time period for all duopolies. Each cell indicates how much revenue the algorithm in the corresponding row was able to earn against the algorithm in the corresponding column on average per time period. Green (red) indicates if the amount was higher (lower) than that of the corresponding opponent.}
\label{tab:duo_revenues}
\end{table}
The last column contains the row-wise averages, which indicate how much revenue each algorithm makes on average overall. Similarly, the final row contains column-wise averages, which indicate how much revenue other algorithms were able to make against the corresponding algorithm (thus, e.g., \logit makes on average 234 per time period, while the other competitors make on average 218 when competing with \logit). The table confirms the observation from Figure \ref{fig:rr_realization} that the market between \wls and \logit is on average the smallest ($153+96=249$ revenue per time period). In fact, \wls proves to be very hard to generate revenue against, since on average competitors earn only $175$ per time period when competing with \wls. Meanwhile, \wls is able to generate substantial profits with an average revenue per time period of $240$, and as a result \wls was able to beat all other competitors based on  revenue earned. This is remarkable, as \wls performs poorly in the oligopoly, as was discussed in the previous section.  

Furthermore, we observe that the results are very mixed and that there is no competitor that loses against all other competitors. Even the worst performer, \ml, is able to defeat other competitors, namely, \logit and \greedy, which is remarkable as both perform very well overall. In addition, based on Table \ref{tab:duo_revenues} we observe that \greedy has a very steady performance, never earning much more (or much less) than its opponents, as one would expect---it simply follows its opponent's actions, without exploiting the opponent's weakness. The other top performer in the duopoly part, namely \ols, only beats \mabgrid and \ml according to Table \ref{tab:duo_revenues}. Nonetheless, we observe that \ols on average earns the same amount of revenue, namely 240, as \wls.


All in all, Table \ref{tab:duo_revenues} reveals that the performance is very much dependent on the competitor's policy and that some algorithms that perform well in the oligopoly (e.g., \greedy), struggle in the duopolies and vice versa (e.g., \ols). This, once more, confirms the intrinsic complexity of pricing and learning with competition.

\begin{figure}
\centering
\textbf{Heatmap of mean prices and price dispersion}\par\medskip
\includegraphics[width=0.65\textwidth]{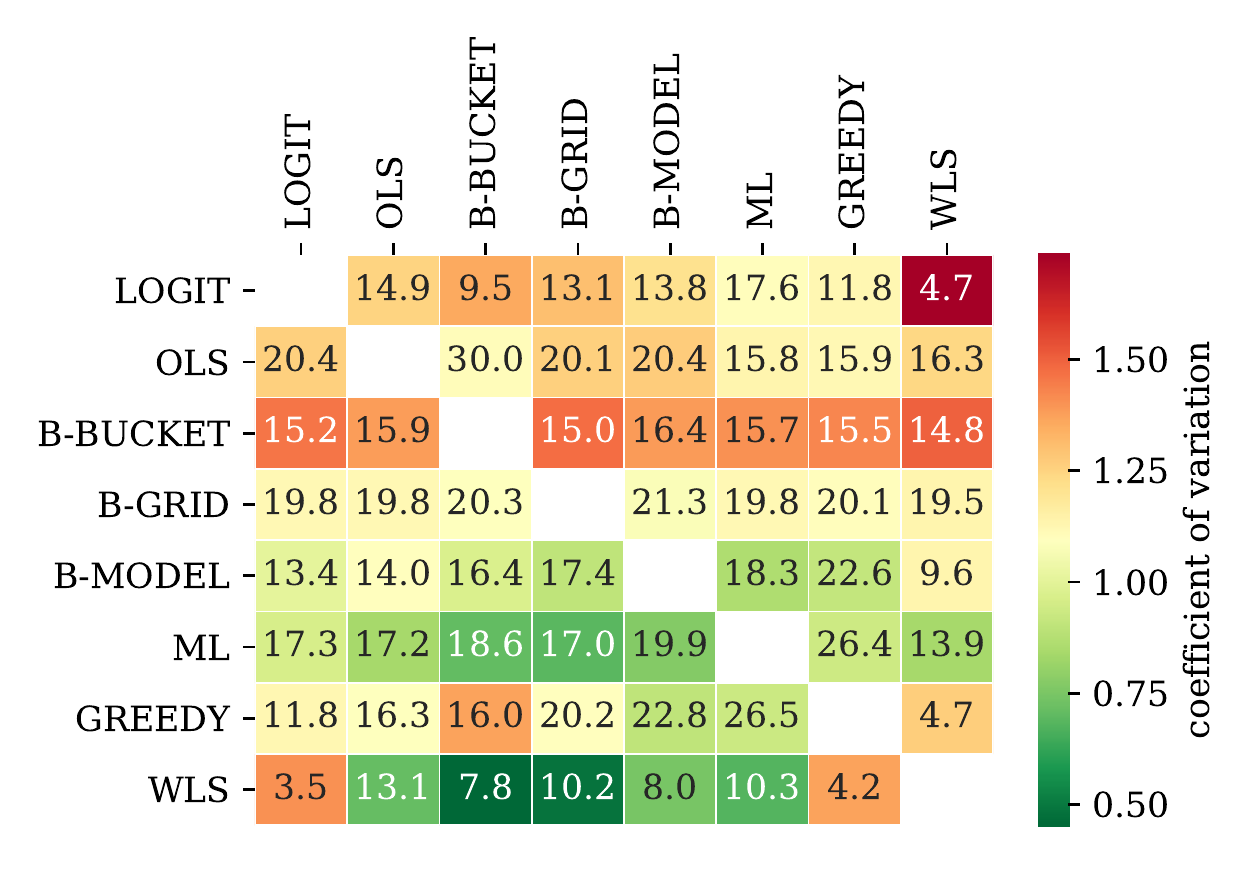}
\caption{The number in each cell indicates the mean price that the algorithm in the corresponding row posted against the algorithm in the corresponding column. The color of each cell pertains to the coefficient of variation (i.e., the ratio of the standard deviation to the mean) and corresponds to the color bar on the right.}
\label{fig:heatmap_prices}
\end{figure}
%
%

Finally, to gain insight in the pricing levels of the competitors in the duopoly competitions, in Figure \ref{fig:heatmap_prices} a heatmap of the mean prices and the coefficient of variation (the ratio of the standard deviation to the mean) is provided. The numerical values correspond to the mean prices posted and the color indicates the amount of dispersion in the prices. 
We observe that \wls consistently has the lowest mean price and that \ols, which performed equally well according to Figure \ref{fig:revbars}, posted on average substantially higher prices than \wls. Thus, \ols is able to remain competitive while maintaining a higher overall price level. Other algorithms, e.g., \mabgrid, also maintain a high mean price level, but are less successful in generating profit. This can well be explained by the fact that the price level of \mabgrid is much less dependent on its competitor, i.e., its mean price posted is always around $20$, whereas \ols varies its pricing level across its competitors, thereby being able to generate more profit. Regarding the dispersion of prices, we observe that \wls, \mabmodels, and \ml are the least experimental in setting prices and that especially \mabbuckets is very experimental, which is in correspondence with Figures \ref{fig:fc_realization} and \ref{fig:rr_realization}.
\section{Discussion}\label{sec:discussion}
In terms of overall performance, \logit has proven to be the most effective algorithm in this competition---it earned most revenue in the oligopoly part and was also competitive in the duopoly part, where it finished in third place, close behind \ols and \wls. Its success can partly be explained by the fact that its demand specification closely resembles the demand function of the scientist segment, which led to the highest revenue per arriving scientist (see Figures \ref{fig:sales_rev_per_arrival} (b) and \ref{fig:r_vs_scientistshare}). However, the fact that \logit was able to identify and exploit the revenue potential of the shopper segment by often pricing low was undoubtedly the key reason for its success. This strategy negatively affected performance when pricing power increased (see Figure \ref{fig:r_vs_loyalshare}), but proved beneficial overall. Arguably, if \greedy would not have had arrangements in place to prevent a `race to the bottom', \logit's revenue might have been substantially lower. This indicates that \logit's low-pricing strategy will not perform well in all circumstances---its dependency on shoppers and scientists makes \logit vulnerable in case the market consists of more competitors that are aggressive on price. In terms of robustness, \mabmodels performs well as it is less reliant on the price-sensitive segments, whilst still earning substantial revenue. 

Regarding the two linear models (\wls and \ols), the difference in their performance between the duopoly and oligopoly parts is striking. In the duopoly setup, \wls and \ols are the two top performers, thereby confirming that the use of linear approximations for demand can give a simple and robust way to model demand for local price changes. The design of \wls, which anticipates its competitor's revenue by maximizing the difference of own and the competitor's revenue, has caused \wls to be very difficult to earn revenue against (see Table \ref{tab:duo_revenues}). On the other hand, \ols ignored competition altogether, which worked surprisingly well in the duopoly competitions and resulted in \ols posting much higher prices than \wls according to Figure \ref{fig:heatmap_prices} (but generating roughly the same amount of revenue). In the oligopoly setup, however, \wls and especially \ols struggled, which indicates that ignoring competition is increasingly harmful as competition increases and that, regarding \wls, explicitly taking into account all competitors' anticipated revenues has proven ineffective. Similarly, the relative performance of \mabgrid, which also ignored competition, was worse in the oligopoly. 

Two of the bandit models, namely \mabgrid and \mabbuckets, performed poorly due to a defect in their designs, which only allowed prices in very crudely discretized price sets and prevented them to set low prices consistently thereby greatly hindering their performance. The algorithms could be improved by making them more adaptive by, e.g., allowing poorly performing arms to be eliminated or adding additional arms close to the current optimal value to allow the algorithm to focus in on profitable prices. It would also be interesting to assess the performance of existing continuous armed-bandit algorithms.

However, the other bandit model, \mabmodels, proved to be competitive and finished third overall. It was designed to cope with different customer behaviors (e.g., bargain hunters and quality seekers, see the appendix for details) by assigning a different demand model to each of its arms. The relative success of the \mabmodels appears to have been its ability to generate above average revenues from each customer segment as, e.g., illustrated in Figure \ref{fig:sales_rev_per_arrival} (b). One of its pitfalls has been its high level of exploration of price points as illustrated in, e.g., Figures \ref{fig:fc_realization} and \ref{fig:boxplot_prices}, especially in comparison to \logit and \greedy. 

The approach of \ml heavily relied on machine learning methods to model the demand characteristics and was designed to cope with non-stationarities, such as changes in the price elasticity over time or changes in the overall demand volume. In doing so, \ml persisted in engaging in exploration cycles, which hindered its performance in the stationary environment that was simulated. This confirms the notion that one should only experiment if the anticipated long-term revenue of doing so outweighs the short-term cost. 

Finally, certainly the most simple strategy, namely \greedy's ad-hoc approach of matching the lowest price in the market, turned out to be very effective in the oligopoly. Generally, it followed whoever was the lowest priced (mostly \logit and \mabmodels) and was thereby able to generate substantial revenue especially from scientists and shoppers. The arrangements that \greedy put in place to prevent downward price spirals were initiated frequently  (see, e.g., Figure \ref{fig:rr_realization}), which could otherwise have led to even lower prices and presumably deteriorating revenues. In the duopoly part, no competitor was able to significantly outperform \greedy, as one would expect, however, \greedy was not able to exploit competitors' weaknesses either, leading to average performance. 
\section{Conclusion and Managerial Insights}\label{sec:conclusion}
This paper presented the results of the Dynamic Pricing Challenge, held on the occasion of the 17\textsuperscript{th} INFORMS Revenue Management and Pricing Section Conference on June 29-30, 2017, at the Centrum Wiskunde \& Informatica, Amsterdam, The Netherlands. The participants of this pricing challenge submitted a wide variety of pricing and learning algorithms of which the numerical performance in a simulated market environment with competition was analyzed. As such, this paper presented a framework in which various paradigms from the field of pricing and learning with competition are analyzed by means of a controlled field experiment. This has allowed us to consider market dynamics that are not analytically tractable or can not be empirically analyzed due to practical complications.

Our analysis has revealed a number of interesting insights both from practical and scientific points of view. First of all, it is shown that relative performance varies substantially across oligopoly and duopoly markets and across different market dynamics, which confirms the intrinsic complexity of pricing and learning in the presence of competition. Most notably, none of the considered algorithms is able to consistently outperform the market---each algorithm meets its Waterloo at some point in the competition. This reveals that algorithm design needs careful consideration and that the structure and dynamics of the market need to be taken into account to determine which algorithm is the best fit. Second, a greedy algorithm that followed the lowest-priced competitor in a tit-for-tat fashion has proved very difficult to outperform. Especially in the oligopolistic markets it was able to attract substantial revenue from price-sensitive customers, whilst---as expected---showing average performance in the duopoly competitions. Third, although the eventual winner was determined by revenue earned, the results reveal that some algorithms are better capable of attracting customers from different segments, thereby being less reliant on one specific segment, and, therefore, being more robust. The winning algorithm, e.g., is predominantly dependent on price-sensitive customers that can easily be targeted by competitors, whilst other competitors earned reasonable revenue from a more loyal customer base. Fourth, the results reveal that ignoring competition is increasingly harmful when competition is more fierce, i.e., when the number of competitors in the market is large and/or price sensitivity of the customers is high. Finally, the analysis reveals that too much exploration can hurt performance significantly. 

Possible extensions to this study that could enhance its generalizability is to impose more complex market dynamics in the simulations, such as temporal dependencies or strategic customer behavior. Although it is appealing (and not too difficult) to do this, we chose not to do so, since it makes it more cumbersome to relate the algorithms and their performance to the dynamics of the market.

%
%
\section{Appendix}\label{sec:appendix}
%
%
\subsection{Competitor Algorithms}
%
%
\subsubsection{Competitor  \mdseries{\logit}}
This competitor models the demand according to a finite mixture logit model, where the mixture is taken over the number of possible customer arrivals. Thus, a probability distribution over the the number of arrivals in a single period is estimated and for each possible number of arrivals, a different multinomial logit model is estimated as well. Each multinomial logit model here, induces a probability distribution over the competitors, i.e., it specifies with which probability an arriving customer purchases from each competitor (including a no-purchase option). In doing so, it is assumed that the utility of buying from competitor $i$ is of the form $a-bp_i$, where $p_i$ is the price posted by competitor $i$ and $a$ and $b$ are assumed to differ across the mixture components. 

In practice, this competitor uses the first 100 time periods to estimate the maximum number of arriving customers in a single time period. This is done by setting a price of 0 for the first period and for each of the following 99 periods of this exploration phase, the price is set as the minimum of the prices observed in the previous period. 
After these 100 periods, an upper bound on the number of arrivals in a single period is taken as the maximum realized demand in a single period multiplied by $(m+1)$, i.e., the number of competitors plus one. Subsequently, an Expectation-Maximization algorithm is used to estimate a probability distribution over the number of arrivals, as well as the parameters of the multinomial logit models. All these parameters are updated every $20$ time periods. 

To optimize prices, in every period the competitors' prices for the period to come are predicted. For this purpose, it is assumed that the sorted prices of the competitors follow a multivariate normal distribution, where the sorted prices are used to mitigate the effect of price symmetries. Subsequently, $1000$ competitor prices are sampled from the multivariate normal distribution and the revenue function is approximated by averaging over these realization. To optimize the price, a crude line search over a discretization of the assumed price space $(0,100)$ is executed and the price with the highest revenue is chosen. 
\subsubsection{Competitor \mdseries{\ols}}
The approach of this competitor to pricing is to favor simplicity. The view is taken that competitors' actions cannot be controlled and that for all intents and purposes, they are random. Thus, they are modeled as an aggregate source of random ``noise'' and the focus is on how the competitor's own price influences demand in this environment. The algorithm is split into an exploration segment and a ``running'' segment. The exploration segment lasts for the first 40 periods and the running segment lasts for the rest of the 960 periods.

In the exploration segment, the algorithm explores the field to ensure sufficient variation in data. In each period, a price is sampled uniformly from the interval $(0,100)$. After the exploration period, the algorithm enters the running segment. In the running segment, the majority of the time consists of estimating a demand curve based only on the competitor's own historical prices and optimizing accordingly. To do so, four linear regression models are fit, taking all combinations of log-transformation of both independent (price) and dependent (demand) variables, and the model with the highest $R^2$ value is chosen (\ols is an acronym for ordinary least squares). Using this model, the price is optimized using a crude line search and, subsequently, a small perturbation is added to the price for further exploration.

Finally, in each period in the running segment there is a 5\% chance of further exploration and a 1\% chance of ``competitive disruption''. Here,  ``competitive disruption'' is an action designed to intentionally confuse competitors who attempt predict competitor prices or who use competitor prices in their model. When this action is initiated the model sets the price to zero in an attempt to confuse competitors via extreme actions. 
\subsubsection{Competitor  \mdseries{\mabgrid}}\label{sec:algD}
\mabgrid is adapted from the $\varepsilon$-greedy multi-armed bandit algorithm \citep{sutton98}. It assumes a bandit framework with ten arms, where the arms pertain to the prices $10, 20, \ldots, 100$ (\mabgrid is an acronym for bandit on a grid). Thus selecting the first arm means posting a price of 10. This algorithm neglects competition and simply keeps track of the average revenue under each arm. With probability $\varepsilon$, an arm is selected randomly, whereas with probability $1-\varepsilon$, the arm that has the highest observed average revenue is selected. The exploration parameter $\varepsilon$ is set to 0.2, so that on average 200 time periods are used for exploration and 800 for exploitation.
\subsubsection{Competitor  \mdseries{\mabbuckets}}
This competitor considers the problem of learning and pricing in a multi-armed bandit framework similar to that of \mabgrid. In doing so, the optimal price is assumed to be contained in the interval $(0, 100]$, which is split into ten intervals of even length, i.e., it is split into price buckets $(0, 10], (10, 20], \ldots , (90, 100]$. Each of these price buckets pertains to one arm and selecting a specific arm means posting a price that is uniformly sampled from the corresponding price bucket (\mabbuckets is an acronym for bandit with buckets).

To incorporate the competitors' prices, it is assumed that the arms' values, i.e., revenues, depend on the prices posted by the other competitors. More precisely, in each time period the competitors' modal price bucket is forecast using exponential smoothing. The modal price bucket is the bucket that is predicted to contain most of the competitors' prices. We assume that the optimal choice of price to offer is dependent on this modal price bucket. 

In practice this works as follows. At each time step, with probability $\varepsilon$ an exploration step is performed in which an arm is selected randomly. Alternatively, with probability $1-\varepsilon$, an exploitation step is undertaken. In this case, the algorithm selects a price from the price bucket with the highest observed average revenue for the predicted modal price bucket. The exploration parameter $\varepsilon$ is set to 0.2, so that on average 200 time periods are used for exploration and 800 for exploitation.
\subsubsection{Competitor  \mdseries{\mabmodels}}
This competitor advocates a bandit formulation of the problem as well, although its design differs conceptually from that of \mabbuckets and \mabgrid. Where the aforementioned two competitors assign prices (or prices buckets) to arms, here, an arm pertains to a demand model (\mabmodels is an acronym for bandit with models). The demand models that constitute the four arms are the following:
\begin{itemize}
\item \textbf{Demand Model 1} (Bargain hunters) assumes that the distribution of customers' willingness to pay (WTP) is normally distributed and that customers select a competitor's price from the subset of prices that fall below their WTP with probability proportional to $\left(\frac{WTP-p_{i}}{WTP}\right)^{b}$, where $p_i$ is the price being offered by competitor $i$ and $b$ is a parameter that influences customers' price sensitivity.
%
High (low) values of $b>1$ ($<1$) capture customer populations that are highly (in)sensitive to prices close to their reserve price. In general this first demand model captures bargain hunters as in all cases customers will tend to choose low prices where possible. 
\item \textbf{Demand Model 2} (Quality seekers) is a variant of the first demand model but the reserve price of a customer is proportional to $\left(1-\frac{WTP-p_{i}}{WTP}\right)^{c}$. This model captures customers who use price as an indicator of quality. The parameter $c$ has a similar interpretation to $b$ above. 
\item \textbf{Demand Model 3} (Cheapest price subset) assumes that each customer sees a different random subset of the available prices. Customers are assumed to select the cheapest price that is visible to them. The subset is assumed to include a random number of options uniformly distributed between $d$ and $e$, which are parameters that can be estimated from the demand data.
\end{itemize}
The fourth arm alone is used for the first 100 time periods with a relatively high exploration rate to provide sufficient data for estimating the parameters $a, b, c$, and $d$ of the three demand models by means of simulated annealing. After 100 time periods the reward vectors are reset and the four-armed bandit assumes control of pricing. Similar to the previous two bandit algorithms, with probability $\varepsilon$ an arm is selected randomly and otherwise the most profitable arm is selected.

Optimal prices are chosen based on a forecast of the competitor price (duopoly) or the profile of competitor prices (oligopoly), where we define the competitor price profile to be an ordered list of competitors' prices. For the oligopoly the competitor price profile is forecast for the next time period using exponential smoothing with trend. In order to estimate the optimal price to charge under each demand model, the algorithm generates a set of potential prices and the projected revenue is evaluated at each price, for the forecast competitor price profile. The price with the highest predicted revenue is assumed to be the best price for this demand model. In an exploitation step, the algorithm selects the arm with the highest predicted revenue and offers the best price for this arm.
\subsubsection{Competitor  \mdseries{\ml}}
The approach of this competitor is to rely on machine learning techniques to predict demand and optimize prices accordingly (\ml is an acronym for machine learning). Much emphasis is put on learning the demand characteristics, as the algorithm dynamically switches back and forth from exploration to exploitation mode over time. In exploration mode, during forty time periods, prices are set according to a cosine function around the mean price level observed to test a variety of price levels and, possibly, confuse competitors. After this learning cycle, demand is modeled using own prices and the competitor prices as covariates by means of a variety of regression models (least-squares, ridge regression, Lasso regression, Bayesian ridge regression, stochastic gradient descent regression, and random forest) and the best model, in terms of demand prediction, is selected through cross-validation. 

Subsequently, the model of choice is used during an exploitation cycle of variable length: the length is sampled uniformly between 70 and 150, however, if the revenue earned deteriorates too fast, then, immediately a new exploration cycle is initiated. The price is optimized by discretizing the price space and computing the revenue for all prices. When a new exploration cycle starts, so either when the exploitation cycle was finished or because the revenue deteriorated significantly, all historical data is disregarded for the benefit of capturing shifts and shocks in the market most adequately. 
\subsubsection{Competitor \mdseries{\greedy}}
This competitor advocates a particularly simple strategy: set the price as the minimum price observed in the previous time period. To avoid a ``race to the bottom'' with another competitor, the following facility is implemented: if the minimum price observed in the previous period is lower than the 10\% percentile of all the prices observed in the last 30 time periods, then the price for the coming period is set as the maximum of this percentile and 5 (i.e., if this 10\% percentile is smaller than 5, the price is set to 5). 
\subsubsection{Competitor \mdseries{\wls}}
The characterizing feature of this competitor is that it aims to maximize own revenue relative to its competitors. More precisely, it attempts to maximize own revenue minus the revenue of the competitor that earns the most revenue. In doing so, it is assumed that demand of competitor $k$ equals $d(p_k,p_{\bar{k}})$, where $p_k$ is the price of competitor $k$, $p_{\bar{k}}$ is the $(m-1)$-vector with the prices of the competitors of $k$, and where the notion of time is suppressed. In addition, it is assumed that $d(\cdot,\cdot)$ is independent of permutations in its second argument, i.e., in the vector $p_{\bar{k}}$.
Thus, this algorithm aims to obtain the price that maximizes own profit compared to the competitors, that is to solve in each time step,
\begin{gather}\label{eq:asbjorn}
  \max_{p_1} \{p_1 d(p_1,p_{\bar{1}}) - \max\{p_k d(p_k,p_{\bar{k}})\mid k = 2,\hdots,m\}\}.
\end{gather}
where the competitors are indexed $1$ to $m$ and \wls is indexed 1. Note that $p_1\in p_{\bar{k}}$ for $k\in\{2,\hdots,m\}$.
 
The demand function is assumed to be of the form $d(x,y) = a+bx+c\sum_{k=1}^{m-1} y_k$ and the parameters $a$, $b$, and $c$ are estimated using weighted least squares (hence the name \wls). To capture different time-dependent aspects of demand, various schemes for the weighting of observations are considered and evaluated based on the Median Absolute Error of their historical demand predictions. The best weighting scheme is used in \eqref{eq:asbjorn} to optimize the price. For this purpose, the price for competitor $k$ in the period to come is predicted based on the median of the historical prices over some window, where the window length is chosen to minimize the Median Absolute Error of historical price predictions. 

Finally, for the purpose of exploration, during the first ten periods prices are randomized to guarantee sufficient variance in the observations to estimate the demand models. In addition, when after these ten periods this competitor's own price is constant for three subsequent periods, the prices are randomized for the next period to induce exploration. 
\clearpage
\fontsize{9pt}{9pt}\selectfont
\bibliographystyle{plainnat}
\bibliography{main}
\normalsize
\end{document}